\documentstyle[aas2pp4]{article}
\def \hb {H$\beta$}
\begin{document}

\title{ON QUASAR MASSES AND QUASAR HOST GALAXIES}
\author{Ari Laor}
\affil{Physics Department, Technion, Haifa 32000, Israel\\
laor@physics.technion.ac.il}

\begin{abstract}
The mass of massive black holes in
quasar cores can be deduced using the typical velocities 
of \hb-emitting clouds in the Broad Line Region (BLR) and the 
size of this region. 
However, this estimate 
depends on various assumptions and is susceptible to
large systematic errors. The \hb-deduced black hole mass in a sample 
of 14 bright quasars 
is found here to correlate with the quasar host galaxy  
luminosity, as determined with the {\em Hubble Space 
Telescope (HST)}. This correlation is similar to
the black hole mass vs. bulge luminosity
correlation found by Magorrian et al. in a sample of 32 nearby
normal galaxies. The similarity of the two correlations is remarkable 
since the two samples 
involve apparently
different types of objects and since the black hole mass estimates
in quasars and in nearby galaxies are based on very different methods. 

This similarity provides a ``calibration'' of the
\hb-deduced black hole mass estimate, suggesting it is accurate to
$\pm 0.5$ on log scale. The similarity of the two correlations also suggests
that quasars reside in otherwise normal galaxies, and that
the luminosity of quasar hosts can be estimated
to $\pm 0.5$~mag based on the quasar continuum luminosity and
the \hb\ line width.
Future imaging observations of additional broad-line active galaxies 
with the {\em HST} are required in order to explore the extent, slope, 
and scatter of the black hole mass vs. host bulge luminosity 
correlation in active galaxies.

\end{abstract}


\section{INTRODUCTION}

Indirect evidence for the existence of massive black holes (MBHs) in 
Active Galactic Nuclei (AGNs) has been growing over the
years (e.g. Rees 1984). However, the most conclusive 
evidence for 
the existence of massive black holes has been recently obtained in
the Milky Way (e.g. Genzel et al. 1997), and in NGC~4258
(Miyoshi et al. 1995), a weakly-active galaxy.
This new evidence is 
based on high spatial resolution observations of stellar and gas kinematics. 
Similar estimates
could not be employed in quasars and bright Seyfert galaxies as the 
stellar kinematics close to the black hole is hopelessly lost behind 
the glare of the active nucleus. A rough estimate of the black
hole mass in AGNs can be obtained based on the size and the typical
velocities in the Broad Line Region (BLR, e.g. Dibai 1981; Wandel \& Yahil 1985;
Joly et al. 1985; Padovani \& Rafanelli 1988; 
Koratkar \& Gaskell 1991). However,
this method is susceptible to various systematic errors, and there is currently 
no independent way to estimate its accuracy. 

Compact massive dark objects, most likely MBHs, were inferred in the
cores of many 
nearby normal
galaxies based on stellar kinematics and the observed light distribution
(see review by Kormendy \& Richstone 1995). In a recent comprehensive study
of the stellar dynamics of a large sample of nearby galaxies
Magorrian et al. (1998) found that a MBH may exist
in the cores of nearly all bulges. They also confirmed  
the strong correlation between the black hole mass and the bulge mass,
consistent with $M_{\rm BH}\sim 0.006M_{\rm bulge}$. If quasars reside
in normal galaxies, then their black hole mass and bulge mass should 
also follow this correlation. It is not possible to explore
this correlation directly 
in quasars since stellar velocity distributions
in the host bulges have not been measured yet. 
However, since there is a strong
correlation between $M_{\rm bulge}$ and $L_{\rm bulge}$ in galaxies
(Faber et al. 1997), one can instead test if quasars follow the $M_{\rm BH}$ 
versus 
$L_{\rm bulge}$ correlation found for normal galaxies.

Rather accurate determinations of quasar host galaxy luminosities
were recently obtained by Bahcall et al. (1997) 
for a representative sample of 20 bright low redshift quasars
using the {\em Hubble 
Space Telescope (HST)}. In this {\em Letter} I 
show that the quasar host galaxy luminosity appears to be significantly 
correlated with the H$\beta$-deduced black hole mass, $M_{\rm BH}({\rm H}\beta)$, 
and that this correlation is very 
similar to the $L_{\rm bulge}$ versus $M_{\rm BH}$ relation 
determined by Magorrian et al. for nearby normal galaxies. 
This similarity provides a ``calibration'' for the
$M_{\rm BH}$ estimates in AGNs. 
The $M_{\rm BH}({\rm H}\beta)$ estimation method,  
its application to the Bahcall et al. sample,
and the correlation analysis results are given in \S 2,
and the implications are discussed in \S 3, together with some
predictions of host luminosities which can be tested with {\em HST} in
the near future.

\section{The $M_{\rm BH}({\rm H}\beta)$ versus $L_{\rm host}$ Correlation}

\subsection{$M_{\rm BH}({\rm H}\beta)$}

The black hole mass can be estimated using the the velocity dispersion of 
the H$\beta$ emitting clouds in the
BLR and the size of this region, together with the assumption that the clouds'
motion are virialized, i.e.,
\begin{equation} 
M_{\rm BH}({\rm H}\beta)=R_{\rm BLR}({\rm H}\beta)v_{\rm BLR}^2/G 
\end{equation}
where
$R_{\rm BLR}({\rm H}\beta)$ is the size of the \hb-emitting region in the
BLR and $v_{\rm BLR}$ is the observed \hb\ velocity dispersion. 

Kaspi et al. (1996) 
find $R_{\rm BLR}({\rm H}\beta)=0.014L_{44}^{1/2}$~pc, where $L_{44}$ is the
$0.1-1\mu m$ luminosity in units of $10^{44}$~erg~s$^{-1}$,
assuming $\Omega_0=1.0, H_0=75$~km~s$^{-1}$~Mpc$^{-1}$. 
This relation is equivalent to 
\begin{equation} 
R_{\rm BLR}({\rm H}\beta)=0.086L_{46}^{1/2}~{\rm pc},
\end{equation}
where $L_{46}$ is the Bolometric luminosity in units
of $10^{46}$~erg~s$^{-1}$ (using $L_{\rm Bol}=3L_{0.1-1\mu m}$, e.g.
Fig.7 in Laor \& Draine 1993), and 
$\Omega_0=1.0, H_0=80$~km~s$^{-1}$~Mpc$^{-1}$ which is used 
throughout the paper. The $R_{\rm BLR}\propto L^{1/2}$
scaling is also indicated by the weak luminosity dependence
of AGN emission line spectra, and is also expected based on
dust sublimation which occurs at 
$R_{\rm dust}\simeq 0.2L_{46}^{1/2}~{\rm pc}$ (Laor \& Draine 1993;
Netzer \& Laor 1993).

Thus, just based on $v_{\rm BLR}$, taken here 
as the observed H$\beta$ FWHM, and $L_{46}$, one obtains
the following estimate for the black hole mass from Eqs.1\& 2,
\begin{equation} 
m_9=0.18\Delta v_{3000}^2 L_{46}^{1/2} ,
\end{equation}
where $m_9\equiv M_{\rm BH}({\rm H}\beta)/10^9M_{\odot}$, and\\
 $\Delta v_{3000}\equiv {\rm H}\beta$~FWHM/3000~km~$^{-1}$. 

\subsection{$L_{\rm host}$}

Quasar hosts have been studied extensively from the ground
(e.g. McLeod \& Rieke 1994a, 1994b; Dunlop et al. 1993). However,
separating out the quasar host galaxy clearly requires a high angular
resolution, and thus measurements with the {\em HST} can
provide the most accurate determination of quasar host properties.

I use the sample of 20 luminous low redshift ($z<0.3$) quasars studied
by Bahcall et al. with the HST Wide Field/Planetary Camera-2 (WFPC2). 
This sample is likely to represent the properties of nearby bright 
quasars. In addition, the large sample size, 
the uniform and detailed reduction, and the detection of all quasar 
hosts, make this sample the best one available for exploring the
$M_{\rm BH}$ versus $L_{\rm host}$ relation in quasars.

Some of the host galaxies morphologies were identified
by Bahcall et al. as elliptical, and for these galaxies
$L_{\rm bulge}\equiv L_{\rm host}$ (taken from the best fit 2-D
model in their Table 5).
Other hosts were identified 
as spirals, or interacting, and the value of
$L_{\rm bulge}$ for these objects, required for a direct comparison
with the Magorrian et al. results, is not available. An estimate
of $L_{\rm bulge}$ for the objects which are best fit by an exponential 
disk is obtained by subtracting
the $7.5\le r\le 15$~kpc annular magnitude (their Table 8) from the
total magnitude (their Table 5), yielding $M_V$(inner host). 
The mean $\Delta M_V$(host$-$inner host) is 0.5 mag, which is 
smaller than $\langle \Delta M_B$(total$-$bulge)$\rangle\sim 1-2$~mag 
for early type spiral galaxies (Simien \& de Vaucouleurs 1986),
suggesting that $M_V$(inner host) overestimates $L_{\rm bulge}$.
Although, $M_V$(host) may be underestimated, since the fit does 
not included a bulge component.

\subsection{The correlations}

Table 1 lists the Bahcall et al. quasars used for the correlation
analysis together with their $z$, $M_V$(bulge) for objects with a
de Vaucouleurs fit, or $M_V$(inner host) for objects with an
exponential disk fit, H$\beta$ FWHM,
bolometric luminosity, $M_{\rm BH}$(\hb) as deduced from Equation 3,
and the host morphology from Bahcall et al.
The H$\beta$ FWHM is obtained from Boroson \& Green (1992) which 
provide high quality and uniformly reduced spectra of all 87 $z\le 0.5$
PG quasars (Schmidt \& Green 1983), of which 14 overlap with the
Bahcall et al. sample. Continuum fluxes are available for all of these 14 PG
quasars in Neugebauer et al. (1987), which provides accurate and
uniformly reduced continuum spectrophotometry for most PG quasars.
The luminosity at 3000\AA\ is converted to $L_{\rm Bol}$ using
$L_{\rm Bol}=8.3\times \nu L_{\nu}$(3000\AA) (see Fig.7
in Laor \& Draine).

There is no uniform data set with the continuum flux and
H$\beta$ FWHM for 5 additional quasars from the Bahcall et al.
sample. Different papers quote parameters that can differ by $>50$\% for a
given object. These objects were therefore not included in
the analysis as they may be subject to significant systematic deviations.

The upper panel in Figure 1 shows $M_V$(bulge/inner host) [hereafter
$M_V$(b/ih)] versus $M_{\rm BH}$
for the 19 Bahcall et al. quasars. Only the 14 quasars
marked with filled squares were used in the analysis.
The Spearman rank order
correlation coefficient is $-0.70$ which has a probability of 0.005 to
occur for unrelated parameters.
A simple least squares fit to the data gives 
\begin{equation} 
M_V({\rm b/ih})=-21.76\pm 0.24-(1.41\pm 0.38)\log m_9 .
\end{equation}
Three quasars which are best fit by an exponential disk, PKS~1302-102,
PG~1307+085, and PG~1444+407 (Bahcall et al. Table 5), may have an elliptical
morphology (Table 1). A least squares fit for the 14 quasars 
using the three quasars de Vaucouleurs fit $M_V({\rm bulge})$
yields the coefficients $-21.85\pm 0.28$ and $-1.18\pm 0.44$.

The middle panel in Fig.1 shows the $M_V$(bulge) vs. $M_{\rm BH}$ relation
for nearby normal galaxies from Magorrian et al. (1998).
The dashed line represents the relation 
\begin{equation} 
M_V({\rm bulge})=-21.40-(2.21\pm 0.28)\log m_9 ,
\end{equation}
as deduced from the
\begin{equation} 
\log (M_{\rm BH}/M_{\odot})=-1.79+(0.96\pm 0.12)\log (M_{\rm bulge}/M_{\odot}), 
\end{equation}
and 
\begin{equation} 
\log (M_{\rm bulge}/M_{\odot})=-1.11+(1.18\pm 0.03)\log (L_{\rm bulge}/L_{\odot}) 
\end{equation}
relations found by Magorrian et al., and the standard relation
$M_V({\rm bulge})=4.83-2.5\log(L_{\rm bulge}/L_{\odot})$.

The quasar correlation is flatter than the Magorrian et al.
correlation ($-1.41$ vs. $-2.21$). This may be partly due to the 
fact that all the quasar hosts at $\log m_9<-0.7$ are disk galaxies, 
where $M_V$(inner host) may overestimate $M_V$(bulge) (\S 2.2).
  
The lower panel in Figure
1 compares directly the distributions of the 32 Magorrian et al.
galaxies and of the 19 quasars in the $M_{\rm BH}$ versus 
$M_V$(b/ih) plane. 
The two distributions overlap surprisingly well.

\section{DISCUSSION}

The overlap of the distributions of quasars and of normal galaxies in the 
$M_{\rm BH}$ versus $M_V$(b/ih) plane is the main result of this
paper. This overlap is remarkable as bright quasars and nearby galaxies are 
apparently different types of objects, and since the $M_{\rm BH}$ estimates 
in quasars and in nearby galaxies are based on very different methods 
(the BLR versus stellar dynamics). 
The overlap is also surprising given the crudeness
of the $M_{\rm BH}$ estimates for both populations. As stressed
by Magorrian et al., their data are fit with a simplified, axisymmetric,
stellar dynamics model, and a more general model may yield $M_{\rm BH}$
which could be off by an order of magnitude,
or may even not require a ``massive dark object'' at all.
The $M_{\rm BH}({\rm H}\beta)$ estimate is even cruder. Large
systematic errors could be induced if the BLR velocity field or
the optical-UV continuum are anisotropic, if the scaling of   
$R_{\rm BLR}({\rm H}\beta)$ with $L$ does not hold in bright quasars,
or if the H$\beta$ dynamics is affected by non gravitational forces
(e.g. radiation pressure, magnetic fields).

The overlap suggests a number of interesting implications.
First, concerning the BLR: 1. The H$\beta$ dynamics are most likely 
dominated by gravity; 
2. the H$\beta$ velocity field and the observed optical-UV emission 
are not likely to be strongly anisotropic, 
and 3. the $R_{\rm BLR}({\rm H}\beta)$ versus $L$ relation
most likely holds in quasars. 
Second, concerning the $M_{\rm BH}({\rm H}\beta)$ estimate; the overlap allows a
``calibration'' of this mass estimate and suggests it is probably 
accurate to within $\pm 0.5$ on log scale. 
Third, concerning quasar hosts; 1. the scatter in the $M_{\rm BH}({\rm H}\beta)$
versus $M_V$(b/ih) correlation suggests that $M_V$(b/ih)
can be estimated to within $\pm 0.5$~mag based on the quasar luminosity
and \hb\ line width. 2. The overlap of the two
distributions suggests that quasar hosts are similar to normal, nearby
galaxies, and thus that $M_V$(b/ih) is generally not strongly affected 
by processes such as a nuclear 
starburst, or distortions due to a tidal interaction. 

This correlation may also be useful for surveys of 
the large scale structure of the universe (2dF, SDSS). Quasars
can be used as bright markers of galaxies out to high $z$ 
whose bulge luminosity
and mass can be deduced from the quasar emission spectra, allowing
studies of clustering as a function of mass.

The correlation found here, 
$ M_V({\rm b/ih})\propto M_{\rm BH}^{-1.4\pm 0.4}$ 
(or $M_{\rm BH}^{-1.2\pm 0.4}$) translates using Eq.7 to
$M_{\rm BH}\propto M^{1.5\pm 0.4}_{\rm b/ih}$ (or 
$M^{1.8\pm 0.6}_{\rm b/ih}$), 
which is steeper than the Magorrian et al. relation
$M_{\rm BH}\propto M_{\rm bulge}$. The slope of the quasar
relation has a relatively large uncertainty due to the small range in
$M_{\rm BH}$ available ($-1.16\le \log m_9\le 0.17$), but it is
interesting to note that at the high mass end ($\log m_9> 0.2$)
the Magorrian et al. galaxies appear to follow the quasar relation 
quite well (see Fig.1).  At the low black hole
mass end one has the two best $M_{\rm BH}$
estimates available, in the Galaxy and in NGC~4258, where
$\log m_9=-2.59; -1.44$ (Miyoshi et al. 1995; Genzel et al. 1997)
and $M_V({\rm bulge})=-18.4; -19.13$ (Bahcall \& Soneira 1980;  RC2
catalogue + Simien \& de Vaucouleurs 1986). These galaxies follow the quasar 
relation significantly better than the Magorrian relation
(see Fig.1). Thus, the data in the range $-2.59\le \log m_9\le 1.2$
appears to agree better with the steeper quasar relation.
The quasar relation is also interestingly close to the Haehnelt, Natarajan 
\& Rees (1998) prediction of $M_{\rm BH}\propto M^{5/3}_{\rm halo}$.

There are some objects, such as M~32, which appear to agree better with
the nearby galaxies relation (Fig.1). 
A number of
Seyfert~1 galaxies with $M_{\rm BH}\sim 10^8-10^9M_{\odot}$, 
as deduced by reverberation
mappings (Peterson et al. 1998), are 1-2 mag brighter
than expected based on the quasar relation (Ho 1998). Subtraction
of the AGN light from the host light would bring them closer to the
quasar relation.

The distribution of quasars in the absolute quasar $B$ band magnitude 
$M_B({\rm quasar})$ versus the absolute host $H$ band magnitude
$M_H({\rm host})$
plane appears to be bounded such that $M_B({\rm quasar})\le M_H({\rm host})$
(e.g. McLeod \& Rieke 1995, their figure 6). McLeod (1998) suggested that 
the reason 
for this bound is that objects where
$M_B({\rm quasar})=M_H({\rm host})$ ``have a maximum allowed black hole mass for their
galaxy mass and that the black hole is accreting at the Eddington
rate.'' This idea is broadly consistent with the correlation found here.
For example, a quasar with $M_H({\rm host})=-25$~mag 
typically has $M_V({\rm b/ih})\simeq -21.3$~mag (using 
$\langle V-H \rangle=3.7$~mag for our 14 PG quasar hosts,
with $M_H({\rm host})$ from McLeod \& Rieke 1994b). 
Equation 4 then gives $\log m_9\simeq -0.33$ (or $-0.47$). A magnitude of
$M_B({\rm quasar})=M_H({\rm host})=-25$~mag 
translates to $\log \nu L_{\nu}(4400$\AA)=45.55,
and $\log L_{\rm Bol}\simeq 46.5$, which corresponds to
0.44 or 0.75 of $L_{\rm Eddington}$. 
The above estimates are rather rough
since the $M_H({\rm host})$ versus $M_V({\rm b/ih})$ correlation
has a significant scatter.

How can the analysis presented here be improved? The crude 
``inner host'' estimate for the bulge luminosity, used here for disk galaxies, 
can be improved by fitting a disk+bulge model to the {\em HST} 
images. This may be feasible for early type spiral hosts where
the typical effective radius of the bulge is $r_e\simeq 1.4$~kpc 
(Simien \& de Vaucouleurs), or $\sim 0''.5$ for the Bahcall et al. 
$z\sim 0.2$ quasars. However, it will not be feasible for late type spiral
hosts, where $r_e\simeq 0.3$~kpc, and a sample of lower $z$ AGNs
will be required. One also needs to measure 
the H$\beta$ FWHM and luminosity simultaneously to guard against 
variability, and to obtain a more
accurate estimate of the ionizing luminosity to use in the
$R_{\rm BLR}({\rm H}\beta)$ relation (Eq.2). Obscuration effects are 
well established
in AGNs, and these may increase the scatter, if not accounted for.
In particular, objects with a very narrow H$\beta$ line may have their
BLR partly obscured, or may be strongly dominated by emission from
the narrow line region. Using the variable H$\beta$ component
profile can overcome such biases.

Future observations with {\em HST} can address the following questions:
1. Does the black hole mass vs. host bulge luminosity correlation 
extends to quasars with higher and lower black hole masses? 
2. Can the scatter in the correlation be
reduced with a more careful analysis, or is it intrinsically 
large, as suggested for galaxies? and 3. Does 
$M_{\rm BH}\propto M_{\rm bulge}^{1.5-1.8}$, as suggested here, or
is $M_{\rm BH}\propto M_{\rm bulge}$
as suggested by Magorrian et al.?

The PG quasars sample may be particularly useful for such future explorations
with {\em HST} since a high S/N homogeneous spectroscopic data base 
is already available from Neugebauer et al. 
and Boroson \& Green. Using this data set and Eqs.3 \& 4 one can
predict that of the 87 $z<0.5$ PG quasars, some of the lowest luminousity hosts 
should be 
found in PG~1244+026, PG~1404+226, and PG~1448+273
(predicted $M_V({\rm b/ih})$=$-18.4$ to $-19.3$), 
while some of the highest 
luminousity hosts should be found in PG~1704+608, PG~1425+267, 
and PG~2308+098 ($-22.0$ to $-22.3$). One can also predict
that the hosts of PG~2304+042 and PG~2209+184 should be 2-3 magnitudes
brighter than the hosts of PG~1244+026 and PG~1448+273 respectively,
although the former and later quasars have, respectively, similar
luminosities. 

\acknowledgments

This work was supported by the fund for the promotion of research
at the Technion. Many thanks to Avi Loeb, Dani Maoz and Don Schneider
for very helpful comments.

\small
\begin{table}
\caption{QUASAR SAMPLE} 
\begin{tabular}{lccrccl}
\tableline \tableline 
Object & $z$ & $M_V^a$ & H$\beta^b$ & $L_{\rm Bol}^c$ & $M_{\rm BH}^d$
& $T^e$ \\
\tableline
PG 0052+251  & 0.155 & $-$20.89  &  5.20   &   46.02   &    8.74 & Sb\\
PHL 909$^f$ & 0.171 & $-$21.48  & 11.0   &   46.37   &    9.57 & E4 \\
NAB 0205+02$^f$ & 0.155  &$-$19.33  &  1.05 &  46.41  & 7.55 & S0? \\
PH 0923+201 & 0.190 & $-$21.48  &  7.61  &    46.06   & 9.09 & E1 \\
PG 0953+414 & 0.239  &$-$20.29  &  3.13  &    46.39   & 8.49 & ? \\
PKS 1004+130 & 0.240 & $-$22.48  &  6.30  &    46.39  & 9.10 & E2 \\
PG 1012+008 & 0.185 & $-$19.91  &  2.64  &    45.85   & 8.07 & Int.\\
HE 1029$-$140$^f$ & 0.086  &$-$20.98  &  7.50 & 46.49 & 9.30 & E1 \\
PG 1116+215 & 0.177 & $-$21.88  &  2.92  &    46.38   & 8.42 & E2 \\
PG 1202+281 & 0.165 & $-$20.98  &  5.05   &   45.51   & 8.46 & E1 \\
3C 273 & 0.158 & $-$22.58  &  3.52   &   46.96   &    8.87   & E4 \\
PKS 1302$-$102 & 0.286 & $-$21.15  &  3.40  &   46.77 & 8.75 & E4? \\
PG 1307+085 & 0.155 & $-$20.51 &   2.36  &    45.99  &  8.04 & E1? \\
PG 1309+355 & 0.184 & $-$21.23  &  2.94  &    45.83  &  8.15 & Sab \\
PG 1402+261 & 0.164 & $-$19.95  &  1.91  &    45.96  &  7.84 & SBb \\
PG 1444+407 & 0.267 & $-$20.49  &  2.48  &    46.13  &  8.16 & E1? \\
3C 323.1 & 0.266  &$-$21.48  &  7.03  &    46.34   &    9.17 & E3? \\
PKS 2135$-$147$^f$ & 0.200  &$-$21.58  & 5.50 & 46.61&  9.09 & E1 \\
PKS 2349$-$014$^f$ & 0.173 & $-$22.58  & 5.50 & 46.57&  9.07 & Int.\\
\tableline 
\normalsize
\end{tabular}
$^a$ Host absolute magnitude from Bahcall et al. (1997, Table 5), 
calculated for $\Omega_0=1.0, H_0=80$~km~s$^{-1}$~Mpc$^{-1}$.\\
$^b$ H$\beta$ FWHM in units of 1000~km~s$^{-1}$ from Boroson \& Green (1992).\\
$^c$ Log Bolometric luminosity in erg~s$^{-1}$, based on $f_{\nu}$ at 
rest frame 3000\AA\ from Neugebauer et al. (1987).\\
$^d$ Log of black hole mass in units of $M_{\odot}$ (see Eq.3).\\
$^e$ Host morphology from Bahcall et al.\\
$^f$ Object not included in the correlation analysis (see text).\\
\end{table}
%\newpage

\onecolumn

\newpage
\begin{figure}
\plotone{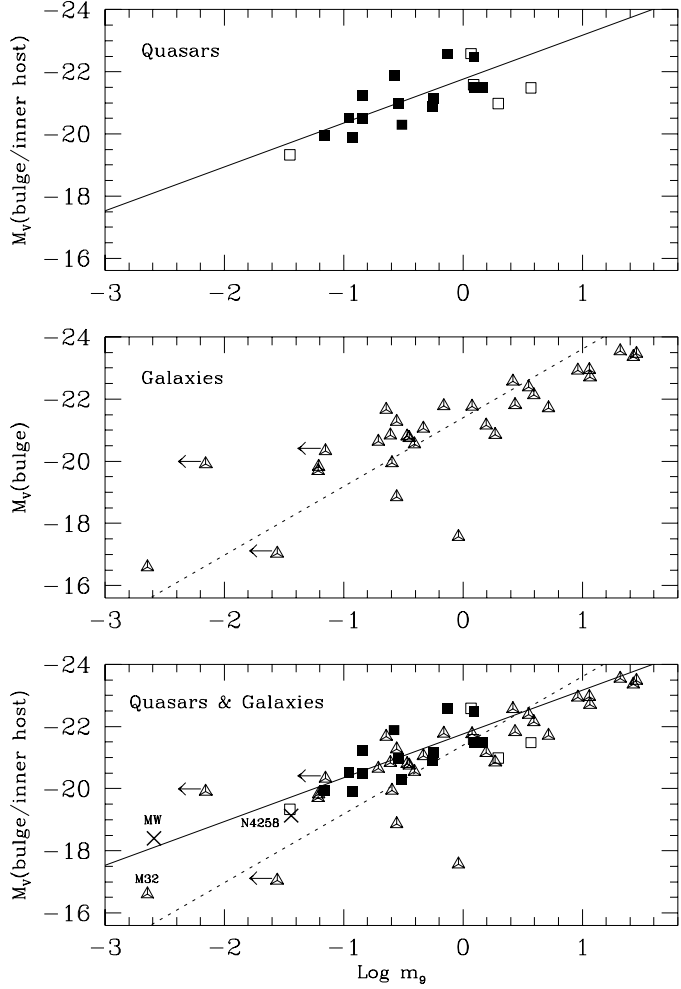}
\caption{Comparison of the correlations for quasars and for nearby
galaxies. Upper panel: the $M_V$(bulge/inner host) versus $M_{\rm BH}$ 
correlation for quasars. The solid line is a least squares fit.
Open squares represent objects which were not included in the fit
(see text). Middle panel: the $M_V(\rm bulge)$ versus $M_{\rm BH}$ 
relation obtained by Magorrian et al. for nearby normal galaxies. 
Lower panel: the two data sets overlaid. The two distributions
overlap surprisingly well. The 
quasar relation is also consistent with the nearby galaxies distribution
at $\log m_9>0.2$, and the positions of the Milky Way and NGC~4258.}
\end{figure}

\end{document}